\documentclass[draft]{agujournal2019}

\usepackage{anyfontsize} 
\usepackage{amsmath,amssymb}
\usepackage{siunitx}

\journalname{Geophysical Research Letters}

\def\bx{\ensuremath{\mathbf x}}
\def\bv{\ensuremath{\mathbf v}}
\def\bu{\ensuremath{\mathbf u}}

\renewcommand{\d}[1]{\ensuremath{\operatorname{d}\!{#1}}}
\newcommand{\D}[1]{\ensuremath{\operatorname{D}\!{#1}}}
\def\pct{\%}

\begin{document}

\title{Clustering of marine-debris-
and \emph{Sargassum}-like drifters explained by inertial particle
dynamics}

\authors{P.\ Miron\affil{1}, M.J.\ Olascoaga\affil{2}, F.J.\
Beron-Vera\affil{1},  N.F.\ Putman\affil{3}, J.\
Tri\~{n}anes\affil{4}\affil{5}\affil{6}, R.\ Lumpkin\affil{4} and
G.J.\ Goni\affil{4}}

\affiliation{1}{Department of Atmospheric Sciences, RSMAS, University
of Miami, Miami, Florida}  \affiliation{2}{Department of Ocean
Sciences, RSMAS, University of Miami, Miami, Florida} \affiliation{3}{LGL
Ecological Research Associates, Inc., Bryan, Texas}
\affiliation{4}{Atlantic Oceanographic and Meteorological Laboratory,
NOAA, Miami, Florida} \affiliation{5}{Cooperative Institute for
Marine \& Atmospheric Studies, University of Miami, Miami, Florida}
\affiliation{6}{CRETUS, Universidade de Santiago de Compostela,
Santiago, Spain}

\correspondingauthor{P. Miron}{pmiron@miami.edu.}

\begin{keypoints}
   \item Debris- and
   \emph{Sargassum}-like oceanographic drifters are observed to
   cluster according to design type.
	\item The clustering is explained as a result of
   inertial effects using a novel Maxey--Riley theory.
	\item The results have implications for understanding
	the movement and distribution of objects such as marine
	debris and pelagic \textit{Sargassum}.
\end{keypoints}

\begin{abstract}
Drifters designed to mimic floating
marine debris and small patches of pelagic \emph{Sargassum} were
satellite tracked in four regions across the North Atlantic. Though
subjected to the same initial conditions at each site, the tracks
of different drifters quickly diverged after deployment. We explain
the clustering of drifter types using a recent Maxey-Riley theory
for surface ocean inertial particle dynamics applied on multidata-based
mesoscale ocean currents and winds from reanalysis. Simulated
trajectories of objects at the air-sea interface are significantly
improved when represented as inertial (accounting for buoyancy and
size), rather than as perfectly Lagrangian (fluid following)
particles. Separation distances between simulated and observed
trajectories were substantially smaller for debris-like drifters
than for \emph{Sargassum}-like drifters, suggesting that additional
consideration of its physical properties relative to fluid velocities
may be useful.  Our findings can be applied to model variability
in movements and distribution of diverse objects floating at the
ocean surface.
\end{abstract}

\section*{Plain Language Summary}

Predicting the fate of floating matter
requires one to recognize that they respond differently than
Lagrangian (i.e., infinitesimally small, neutrally buoyant) particles
to the action of surface currents and winds.  Indeed, the Maxey--Riley
equation of fluid mechanics---a Newton-second-law-type ordinary
differential equation---shows that even the motion of seemingly
small neutrally buoyant particles immersed in a fluid in motion can
substantively deviate from that of Lagrangian particles.  The
Maxey--Riley equation has been recently extended to account for the
combined effects of ocean current and wind drag on finite-size
particles floating at the ocean surface.  We show here that the
paths of  drifters that
mimic marine debris and small \emph{Sargassum} patches cluster
according to their inertial characteristics consistent with the
Maxey--Riley theory .

\section{Introduction}

Floating matter of a very diverse nature ranging from microplastics
\cite{Cozar2014} to larger objects \cite{Maximenko2019, vanSebille2020}
is commonly found throughout the oceans.  The monitoring and forecast
of their trajectories are key to improving the efficiency of marine
debris \cite{Morrison2019} and \emph{Sargassum} \cite{Langin2018}
removal efforts, search-and-rescue operations of different types
\cite{Breivik2013}, and marine safety \cite{Hong2017}.  However,
forecasting the trajectories of floating matter is challenging due
to a number of forcing agents controlling its motion.

Indeed, the well-established fluid mechanics' Maxey--Riley equation
\cite{Maxey1983} dictates from first principles that a finite-size
particle immersed in the flow of a fluid with possible different
density---i.e., an \emph{inertial} particle---will be accelerated
by the undisturbed fluid's flow force and added mass force (resulting
from part of the fluid moving with the particle), while its trajectory
may be deflected by shear-induced lift and Coriolis (in a geophysical
fluid) forces, and slowed down by the drag force (due to the fluid's
viscosity).  The effects of these forces prevent an inertial particle
from adapting its velocity to that of the carrying fluid.  In other
words, inertial particle motion can be quite different than Lagrangian
(i.e., fluid) particle motion \cite{Cartwright2010}.

In this letter we report on the analysis of trajectories produced
by custom-made undrogued surface drifters designed to
mimic floating marine debris of varied sizes and shapes and small
patches of pelagic \emph{Sargassum}.  These special
drifters were deployed together with conventional drifters at several
locations along the path of the 2018 PIRATA (Prediction and Research
Moored Array in the Tropical Atlantic) Northeast Extension cruise
to study how marine debris and \emph{Sargassum} move
under different ocean and wind conditions \cite{Duffy2019} and
assess inertial effects on their drift. The
motion of the drifters is investigated using a Maxey--Riley equation
proposed by \cite{Beron-Vera2019}---henceforth referred to as \emph{BOM
equation}---as a first-principle-based alternative to ad-hoc approaches
commonly taken in oceanography to simulate the influence of ocean
currents and winds on the drift of floating matter \cite{Allshouse2017,
Trinanes2016}.  This work extends the scope of the initial, successful
testing of the BOM equation \cite{Olascoaga2020} by considering various
drifters of the same design type per deployment, which enabled to
observe clustering of paths dominated by inertial effects. It
also considers longer trajectories, which, sampling a wider range
of ocean current and wind conditions, enabled a much more stringent
test of the importance of inertial effects.

\section{Methods}

\begin{figure}[t!]
  \centering%
  \includegraphics[width=\textwidth]{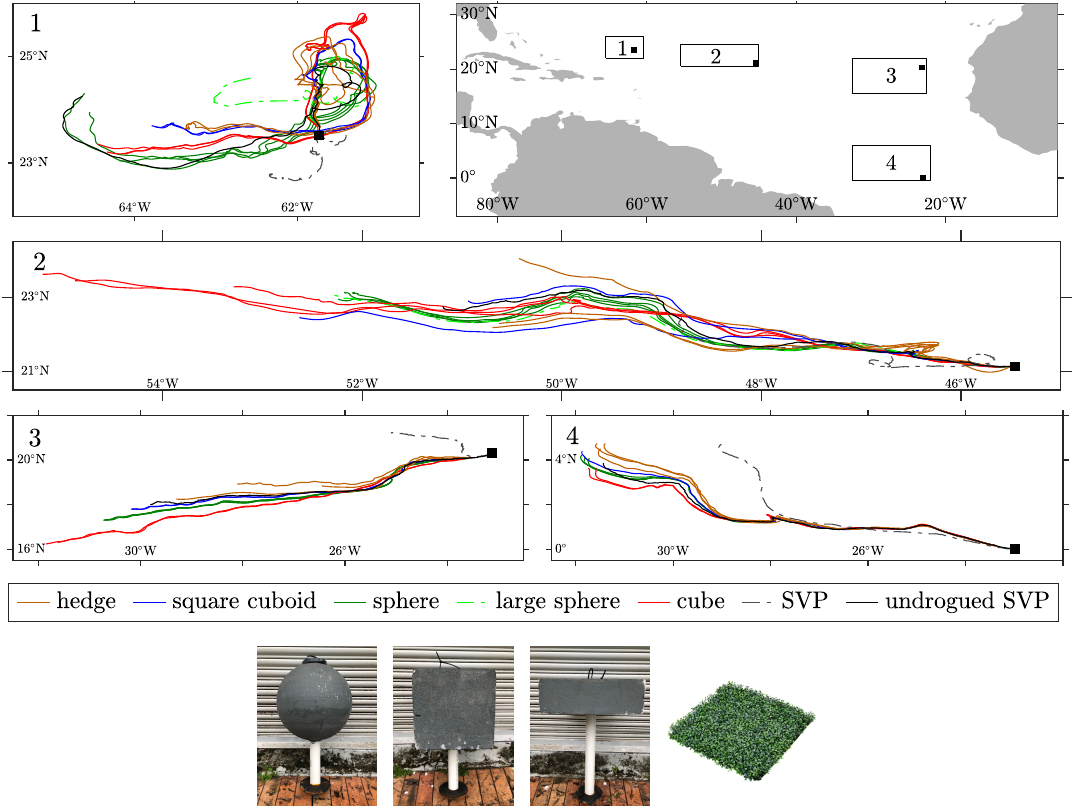}%
  \caption{One-month-long trajectories of the various drifters
  deployed along the PIRATA Northeast Extension cruise in the North
  Atlantic on 11 (1), 14 (2), 20 (3), and 28 (4) March 2018.
  Trajectories produced by debris-like drifters (small
  spheres, large spheres, cubes, and  square
  cuboids) and \emph{Sargassum}-like drifters (hedges), which are
  represented as solid-green, dashed-green, solid-red, solid-blue,
  and solid-brown curves, respectively.  Dashed- and solid-black
  curves are trajectories of drogued and undrogued SVP drifters,
  respectively. The special drifters (excluding the large
  sphere) are displayed at the bottom.}
  \label{fig:dep}%
\end{figure}

The trajectories of the drifters that are the focus of this letter
are depicted in Figure~\ref{fig:dep}. These trajectories were
produced by 47 drifters specially designed at NOAA's Atlantic
Oceanographic and Meteorological Laboratory.  Five types of special
drifters were built. The four debris-like
drifters were comprised of a main body, made of polystyrene foam,
and a short-weighted tube at the bottom.  Two kinds of main bodies
were considered: spherical (of two sizes) and cuboidal (a cube and
a square cuboid).  The volume of the tube was too small to significantly
contribute to the buoyancy of the drifter.  Rather, it ensured that
the trace GPS (Global Positioning System) trackers were maintained
above the sea level at all times, transmitting positions every 6
h. The fifth special drifter type consisted of
an artificial boxwood hedge designed to mimic a small patch
of pelagic \emph{Sargassum} \cite{Olascoaga2020, Putman2020}. The GPS tracker in this
case was placed inside a small polystyrene foam cone inserted in
the hedge. A total of 15 small spheres, 2 large spheres, 10 cubes,
8 square cuboids, and 12 hedges were produced.  Figure~\ref{fig:dep}
includes reference trajectories produced by additional 8 drifters
from NOAA's Global Drifter Program \cite{Lumpkin2007} with positions
satellite tracked using GPS.  Four of these drifters followed the
conventional Surface Velocity Program (SVP) design with a spherical
surface float and a drogue (holey sock) attached to it and centered
at 15-m depth, which serves to minimize wind slippage and wave-induced
drift \cite{Sybrandy1991}.  The other 4 SVP drifters had no drogue
attached.  Four deployments were carried out on 11, 14, 20, and 28
March 2018, each one involving more than one
special drifter of each class, along with one drogued and one
undrogued SVP drifter.  The differences among drifters
used in these experiments ensured that the studies could be done
for a variety of buoyancy and drifter aerial exposure to winds.

The redundancy of special drifter types per deployment enabled a
classification of the motion of the drifters as a function of
inertial characteristics. This classification was carried out using the KMeans
clustering algorithm. Given a number of points distributed in some
space, the algorithm regroups the points into $n$ clusters that
minimizes the square of the pairwise distance between points of a
cluster (also known as the within-cluster sum-of-squares criterion).
The parameter $n$ was set to the a priori known number of distinct
types of special drifters included in each deployment.  Essentially
the same results are obtained using unsupervised techniques such
as DBSCAN (Density-Based Spatial Clustering of Applications with
Noise) \cite{Ester1996}, which adds confidence to the results.

Further insight into the motion of the drifters was attained by
applying the BOM equation, which is formulated as follows.  Consider a
small spherical particle of radius $a$ characterized by negligible
air-to-particle density ratio and finite water-to-particle density
ratio $\delta \ge 1$, loosely referred to as \emph{buoyancy}, so
reserve volume is well approximated by $1 - \delta^{-1} \in [0,1)$
\cite{Olascoaga2020}. If the particle is not spherical, we treat
it as if it was so, but with an effective radius
$a\smash{\sqrt{3a_\mathrm{v}/(a_\mathrm{n} + a_\mathrm{s})}}$, where
$a$ is the average of $a_\mathrm{n}$, $a_\mathrm{s}$, and $a_\mathrm{v}$,
which are the radii of the sphere with equivalent projected area,
surface area, and equivalent volume, respectively.  Let $\bx =
(x,y)$ denote position on the surface of the Earth, where $x =
(\lambda-\lambda_0)a_\odot\cos\vartheta_0$ and $y =
(\vartheta-\vartheta_0)a_\odot$ are northward and eastward local
curvilinear coordinates, respectively, measured from
$(\lambda_0,\vartheta_0)$, with $\lambda$ (resp., $\vartheta$)
longitude (resp., latitude).  Here $a_\odot$ is the Earth's mean
radius and $\gamma_\odot := \sec\vartheta_0\cos\vartheta$ is a
geometric factor (due to the planet's curvature) needed to compute
distance, i.e.,  $\gamma_\odot^2\d{x}^2 + \d{y}^2 =: \d{\bx}^\top
m_\odot^2 \d{\bx}$ gives the arc length square ($m_\odot$ is the
metric matrix with $\gamma_\odot$ in the upper-left entry and 1 in
the lower-right entry).  The BOM equation provides a motion law for the
particle in the form of a second-order ordinary differential equation
given by (the overdot denotes time derivative)
\begin{equation}
  \dot\bv_\mathrm{p} + \left(f + \tau_\odot v_\mathrm{p}^x +
\tfrac{1}{3}R\omega\right)
  \bv_\mathrm{p}^\perp + \frac{\bv_\mathrm{p}}{\tau} = R\frac{\D{\bv}}{\D{t}} +
  R\left(f + \tau_\odot v^x + \tfrac{1}{3}\omega\right)\bv^\perp +
  \frac{\bu}{\tau}, \label{eq:MR}
\end{equation}
where $\bv_\mathrm{p} = m_\odot \dot\bx$ is the particle's
velocity,
\begin{equation}
  \bu := (1-\alpha)\bv + \alpha \bv_\mathrm{a}, 
  \label{eq:u}
\end{equation}
and $\perp$ represents a $+\frac{\pi}{2}$ rotation.

Time-and/or-position-dependent quantities in \eqref{eq:MR}--\eqref{eq:u}
are: the (horizontal) velocity of the water, $\bv = (v^x,v^y)$,
where $v^x$ (resp., $v^y$) is zonal (resp., meridional); its material
(water-particle-following) derivative, $\smash{\frac{\D{}}{\D{t}}}\bv
= \partial_t \bv + \smash{\gamma_\odot^{-1}(\partial_x \bv)}v^x +
(\partial_y \bv)v^y$; the water's vorticity, $\omega =
\gamma_\odot^{-1}\partial_xv^y - \partial_yv^x + \tau_\odot v^x$,
where $\tau_\odot := \smash{a_\odot^{-1}}\tan\vartheta$; the velocity
of the air, $\bv_\mathrm{a}$; and the Coriolis frequency, $f =
2\Omega\sin\vartheta$, where $\Omega$ is the planet's angular
velocity magnitude.  

Primary BOM equation's parameters $a$ and $\delta$ determine secondary
parameters $\alpha$, $R$, and $\tau$ as follows: $\alpha :=
\gamma\Psi/(1 + (1 - \gamma)\Psi) \in [0,1)$, which makes the convex
combination \eqref{eq:u} a weighted average of water and air
velocities ($\gamma \approx 0.0167$ is the air-to-water viscosity
ratio); $R : = (1 - \frac{1}{2}\Phi)/(1 - \frac{1}{6}\Phi) \in
[0,1)$; and $\tau := \smash{\frac{a^2\rho}{3\mu}}\cdot
(1-\frac{1}{6}\Phi)/(1 + (1 - \gamma)\Psi)\delta^4 > 0$, which
measures the \emph{inertial response time} of the medium to the
particle ($\rho$ is the assumed constant water density and $\mu$
the water dynamic viscosity).  Here $\Phi :=
\smash{\frac{\mathrm{i}\sqrt{3}}{2}}(\varphi^{-1} - \varphi) -
\frac{1}{2}(\varphi^{-1} + \varphi) + 1 \in [0,2)$ is the fraction
of emerged particle piece's height, where $\varphi^3 :=
\mathrm{i}\smash{\sqrt{1 - (2\delta^{-1} - 1)^2}} + 2\delta^{-1} -
1$, and $\Psi := \pi^{-1}\cos^{-1}(1 - \Phi) - \pi^{-1}(1 - \Phi)
\smash{\sqrt{1 - (1 - \Phi)^2}} \in [0,1)$, which gives the fraction
of emerged particle's projected (in the flow direction) area.

Details of the derivation of the BOM equation are deferred to
\citeA{Beron-Vera2019} (cf.\ also \citeA{Olascoaga2020} for an
evaluation of its range of validity and a closure proposal).  Here
we simply mention that it mainly follows from vertically integrating
the original Maxey--Riley equation, adapted to account for the
effects of the Earth's rotation and curvature, for a particle
floating at an unperturbed air--sea interface.  In particular we
note that \eqref{eq:u},  which plays an outstanding role in short-term
evolution through its dependence on $\delta$ (buoyancy) as we will
show, follows from integrating the (Stokes) drag term. 

In the simulations discussed below, $\bv$ in the BOM equation was taken
as a daily surface velocity synthesis at
$0.25^\circ$ resolution of geostrophic flow derived from multisatellite
altimetry measurements \cite{LeTraon1998} and Ekman drift induced
by wind from reanalysis \cite{Dee2011}, combined to minimize
differences with velocities of drogued SVP drifters.  Consistent
with the geostrophic component of this $\bv$ representation, we
restricted our simulations to deployments 1--3, which lie sufficiently
away from the equator (cf.\ Figure~\ref{fig:dep}).  In turn,
$\bv_\mathrm{a}$ was taken as the same reanalyzed wind involved in
the surface velocity synthesis. More precisely, $\bv_\mathrm{a}$
is set to half the 10-m height reanalyzed wind \cite{Hsu1994}. These
multidata-based mesoscale ocean currents and winds from reanalysis
were shown capable of representing reality quite accurately in our
first implementation of the BOM equation \cite{Olascoaga2020}.  The
specific values taken by the various (primary and secondary) inertia
parameters for each of the special drifter types are shown in Table
\ref{tab:par}. These represent mean values (the weight of
drifters of the same type typically vary about 1.5\pct), assuming
an average climatological seawater density value of
\protect\SI{1025}{\kilogram\per\meter\cubed}. Integrations were
carried out using \cite{Dormand1980}
scheme with interpolations (in space and time) done using a cubic
method.

\renewcommand{\arraystretch}{1.1}
\begin{table}[t!]
  \linespread{1}\selectfont{}
  \centering
  \begin{tabular}{l*6c}%
    \hline%
	 &  \multicolumn{6}{c}{Parameter}\\
	 \cline{2-7}%
	 &  \multicolumn{2}{c}{Primary} & &
	 \multicolumn{3}{c}{Secondary}\\
	 \cline{2-3}
	 \cline{5-7}
    Drifter & $a$ [\si{cm}] & $\delta$ & & $\alpha$ & $R$ & $\tau$
[\si{d^{-1}}]\\
    \hline
    Small sphere &  12.4 & 4.0 & & 0.043 & 0.42 &
0.0008\\
    Large sphere &  14.0 & 5.5 & & 0.058 & 0.36 &
0.0031\\
    Cube  & 15.9 & 6.8 & & 0.071 & 0.32 & 0.0002\\
    Square cuboid & 12.9 & 4.5 & & 0.048 &
0.40 &
0.0006\\
    Hedge & 21.5 & 1.2 & & 0.005 & 0.79& 0.0320\\
    \hline%
  \end{tabular}\\%
  \caption{Parameters that characterize as inertial ``particles'' the
  special drifters.}
  \label{tab:par}
\end{table}
\renewcommand{\arraystretch}{1}

Finally, Fr\'echet distance between observed and simulated trajectories
was used to quantitatively assess the skill of the BOM equation. The
Fr\'echet distance is the shortest cord-length between two points
traveling on separate curves, possibly at different speeds
\cite{Rockafellar1998}.  Previous studies \citeA{Olascoaga2020}
have used other measures, such as the Hausdorff distance, which is
the longest of all shortest distances between two curves.  The
conclusions we present here are not sensitive to a particular measure
of curve similarity, as the Frechet and Hausdorff distances produced
indistinguishable results.

\section{Results}

We begin with a qualitative description of the observed trajectories
(Figure~\ref{fig:dep}).  While the special and undrogued drifters
exhibit diverse paths among themselves, they differ quite starkly
from those taken by the drogued drifters.  However, a quite evident
aspect of the special drifter trajectories is a tendency to cluster
according to type.  A single color is used to depict trajectories
of special drifters of the same type in Figure~\ref{fig:dep}.  Note
the relatively small spread of curves of the same color.  This
strongly suggests that inertial effects are operative.

The above qualitative
inference is quantified by constructing a matrix $D$ with $(i,j)$th-entry
given by the distance between drifter pair $(i,j)\in I\times I$,
cumulative from deployment until the end of the first week  ($I$
is the set of drifters in question).  Clearly, $D_{ij} = 0$ for $i
= j$.  Figure~2 shows $D$ for each deployment in Figure~1 with $I$
grouped by special drifter type. Cursory inspection of Figure~2
reveals predominantly low values of $D$ for drifters of the same
type, irrespective of the deployment. This is seen more so by
applying KMeans clustering (Sculley, 2020) \cite{Sculley2010} on
the positions of the special drifters at the end of the first week.
The number of clusters extracted is naturally set as the number of
drifters of a given type included in each deployment, i.e., 4, 4,
4, and 3 for deployments 1, 2, 3, and 4, respectively. To visualize
the clustering results, the drifters that form a cluster are
highlighted by a dashed contour on the $D$ plot in Figure~2 and
identified with a roman numeral. The clusters correspond quite well
with hedges, square cuboids, spheres, and cubes, respectively. The
exception is deployment 1, in which case only cubes are identified
as a well-defined cluster. Note that the clustering is not entirely
determined by buoyancy ($\delta$). Indeed, the cuboids ($\delta=4.5$)
tend to drift closer to the hedges ($\delta=1.2$) than the small
spheres ($\delta=4$), as revealed by a systematically lower cumulative
pairwise distance. Furthermore, the hedges and the cuboids are
grouped into the same cluster (labeled I) in deployment 1. This
suggests that size and also (more likely) shape contribute to
controlling the clustering. Overall this quantitative clustering
analysis supports the qualitative clustering assessment above.

\begin{figure}[t!]
    \centering \includegraphics[width=\textwidth]{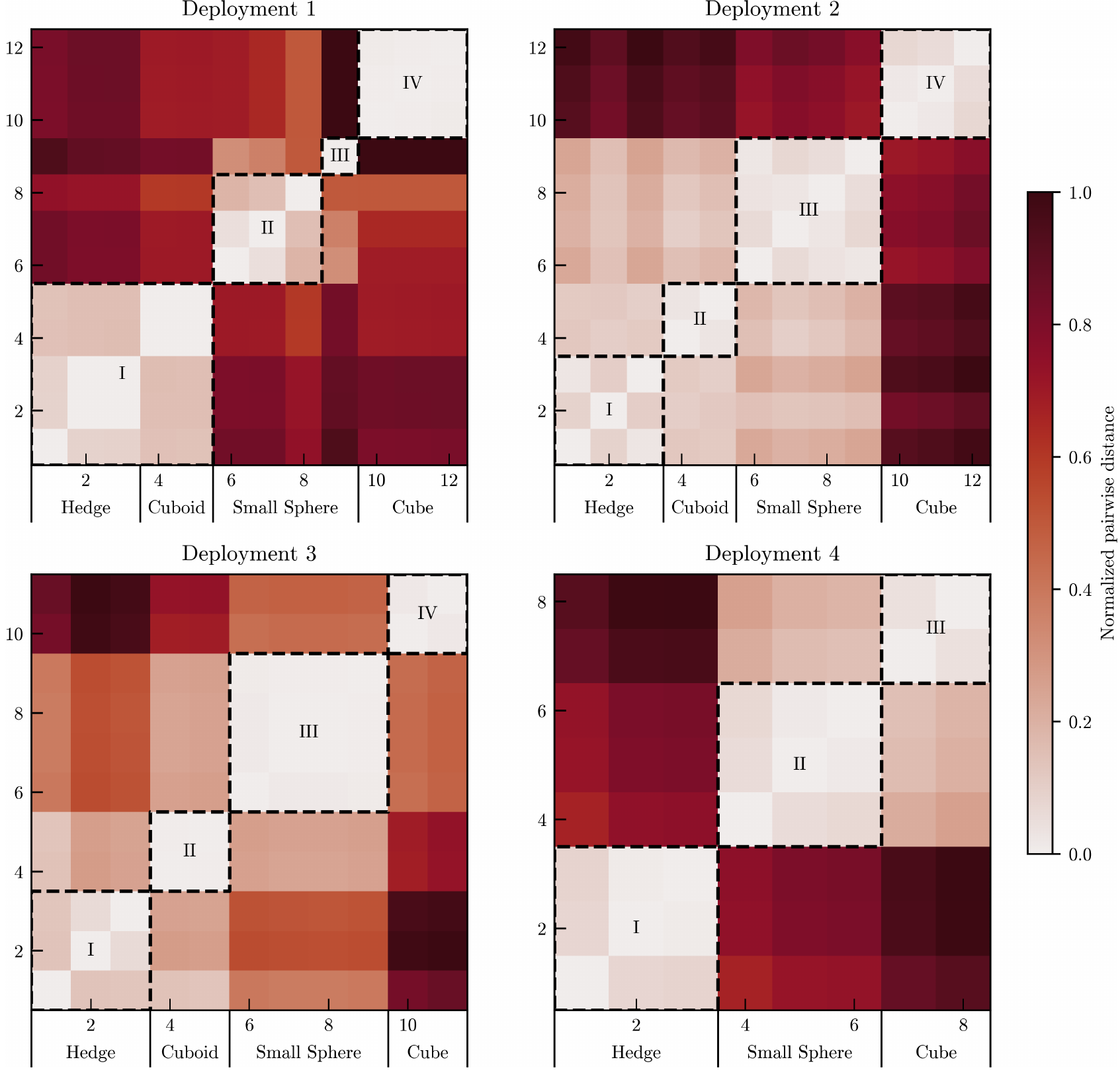}
    \caption{Normalized distance matrix (presented from white to
    red) between pairs of hedges, square cuboids,
    small spheres, and cubes, cumulative
    from deployment out to the end of the first week. Indicated by
    black squares and roman numerals are the clusters revealed
    by KMeans analysis.} 
	 \label{fig:cluster}
\end{figure}

The observed clustering of the special drifters is explained, at
large, by inertial effects as described by the BOM equation.  This follows
from the comparison of special drifter trajectories and trajectories
produced by the BOM equation.  As noted above, excluded from this
assessment are trajectories from deployment 4, which lies too close
to the equator where the surface ocean current representation
considered is not valid.  Our assessment applies to all special
drifters except hedges, which require a different description than
that provided by the BOM equation.  Meant to simulate pelagic \emph{Sargassum}
rafts, a minimal Maxey--Riley model for them would need to include
elastic interactions between gas-filled bladders that keep rafts
afloat.  Such a minimal model has been proposed and applied with
qualitative success \cite{Beron-Vera2020}.  However, the hedges do
not adhere to \emph{Sargassum} rafts' morphology and thus a different
approach is needed.  Indeed, they do not form networks of elastically
interacting inertial particles, but rather water-absorbing objects,
which are not contemplated in the Maxey--Riley framework.  In turn,
differences between trajectories produced by solid special drifters
and simulated counterparts can be attributed to uncertainties around
inertia parameters ($\delta$, primarily), but most likely to those
around the carrying flow determination.

\begin{figure}[t!]
    \centering \includegraphics[width=\textwidth]{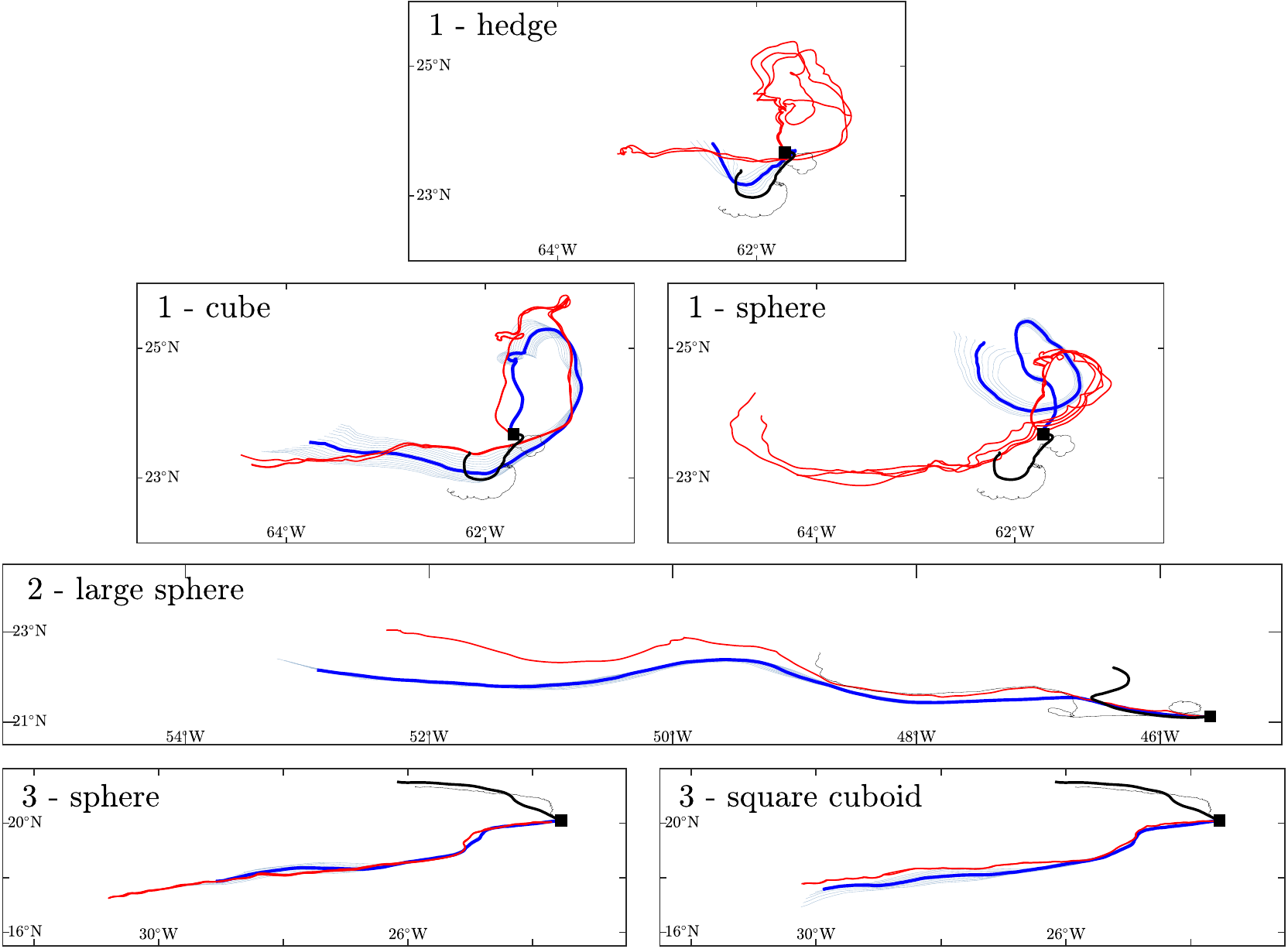}
    \caption{Observed (red) and BOM equation's (blue) one-month-long
    trajectories corresponding to a hedge (top), cube (mid-top-left),
    small sphere (mid-top- and bottom-right), large sphere (middle),
    and square cube (bottom-left).  Light blue curves
    are BOM trajectories with buoyancy allowed to vary 10\pct\,
    around its nominal value. Also shown in each panel is the
    trajectory of one drogued SVP drifter (thin-black) and the
    trajectory produced by the altimetry/wind/drifter synthesis
    that provides a representation for the water velocity component
    of the BOM equation (heavy-black).  Deployment number and special
    drifter are indicated in each panel.} 
	 \label{fig:bom}
\end{figure}

Representative examples of observed and BOM trajectories over a
relatively long period of one month are presented in Figure~\ref{fig:bom}.
In all panels, red curves represent observed special drifter
trajectories, while blue curves are corresponding BOM trajectories
starting from observed initial positions and velocities.  Included
also are the trajectory of a drogued drifter (thin-black) and a
trajectory obtained by integrating $\dot\bx = m_\odot^{-1}\bv$
(recall that $\bv$ is given by an altimetry/wind/drifter synthesis
that is expected to most closely represent drogued drifter velocity).
These examples sample the spectrum of successful, acceptably
successful, not so successful, and mostly unsuccessful simulations.
Beginning on the unsuccessful end, the top panel shows the typical
situation with hedge drift simulation, which is invariably quite
poor.  The mid-top panels provide examples of not-so-successful
simulations, typically seen in deployment 1.  The cube trajectory
(left) is reasonably simulated, while the small sphere trajectory
(right) is not, not even if $\delta$ is allowed to vary 10\pct\,
around its nominal value (light blue).  Note that motion of the
special drifters is quite different than that of the drogued drifter,
which evidently is not completely described by the mesoscale ocean
current synthesis, likely due in part to the smoothness of the
altimeter data used to derive geostrophic currents.  Altogether
ocean and wind representations in this region may be contributing
to hinder the ability of the BOM equation to reproduce observed behavior
\cite{Putman2016}.  The mid-bottom panel shows an example of
acceptably successful simulations, typically seen in deployment 2.
Limitations of the ocean velocity representation, evidenced by its
limited skill in reproducing the drogued SVP trajectory, appears
to be a dominant factor in this case. Buoyancy
variations about the nominal one for the large sphere do not
contribute to reducing the differences with the observed trajectory
 that increase with
time.  Clearly, sensitivity to
initial conditions and accumulated errors are important factors too
in all cases.
Finally, the bottom panels show typical examples of successful
simulations for the case of a square cuboid (right) and a small
sphere (left), mainly happening in deployment 3.  These simulations
reveal the importance of an accurate surface carrying flow
representation (note the very good agreement in this case between
observed and simulated drogued drifter trajectories which contributes
to enhancing the performance of the BOM equation. We note that
the simulated trajectories in all cases are very similar to those
resulting using the $\beta$-plane form of the BOM equation, which
follows by setting, in (1), $\gamma_\odot = 1$, $\tau_\odot = 0$,
and $f = f_0 + \beta y$ while treating $\bx$ as Cartesian.  This
should is anticipated given their limited meridional extent \cite{Ripa1997}.

\begin{figure}[t!]
    \centering \includegraphics[width=\textwidth]{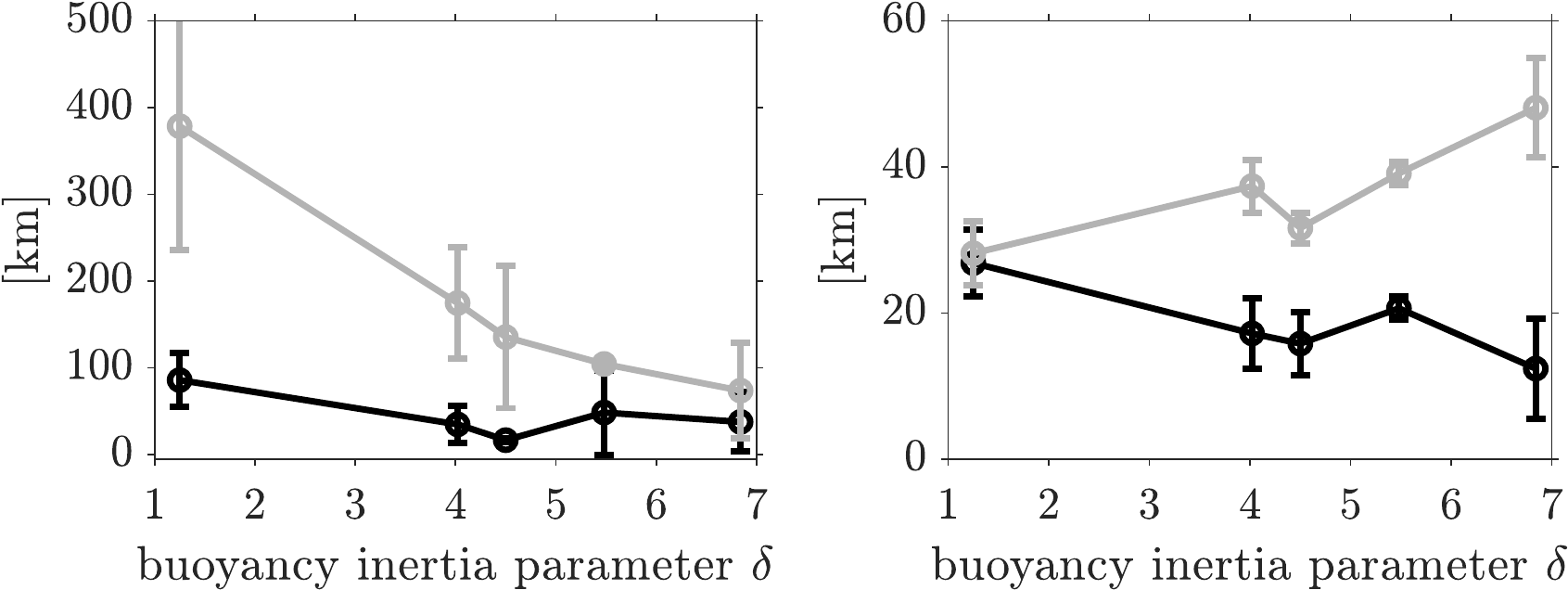}
    \caption{(left panel) As a function of buoyancy
    inertia parameter $\delta$, Fr\'echet distance between observed
    and BOM trajectories after one week (black) and one month (gray).
    Circles are mean values over all distances independent of
    deployment, while error bars are of one standard deviation
    across them.  (right panel) Geodesic distance between observed
    and BOM positions after one day (black) along with that between
    observed position and simulated position based on the
    altimetry/wind/drifter ocean current synthesis (gray).}
    \label{fig:dis}
\end{figure}

Quantitative assessments of the above qualitative assessments of
the BOM equation's skill are presented in Figure~\ref{fig:dis}.  The
left panel shows, as a function of buoyancy, Fr\'echet distance
between observed and BOM trajectories after one week (black) and
one month (gray).  While the variability can be large, the mean
Fr\'echet distances are substantially much smaller for the square
cuboids, cubes, and spheres than for the hedges, consistent with
our qualitative assessments shown above.  Fr\'echet distances
increase with time as can be expected.  The right panel of
Figure~\ref{fig:dis} shows, in a similar manner as in the left
panel, the geodesic distance between observed and BOM position after
one day (black) along with that between observed and simulated
position based on the altimetry/wind/drifter ocean velocity synthesis
(gray).  We have chosen one day as this is a critical time scale
in search-and-rescue operations at sea \cite{Breivik2013}.  Note
that the BOM equation can improve the likelihood of such type of operations
substantively.

We close this letter by noting that BOM trajectories are very similar
to those resulting by integrating $\dot\bx = m_\odot^{-1}\bu$.  The
reason for this is found in the relative smallness of the special
drifters (equivalently, their short inertial response time $\tau$).
Indeed, application of geometric singular perturbation theory
\cite{Fenichel1979, Jones1995} extended to nonautonomous dynamical
systems \cite{Haller2008} on \eqref{eq:MR}--\eqref{eq:u} reveals
\cite{Beron-Vera2019} that the trajectory of a sufficiently small
inertial particle converges in the long run on an attracting
\emph{slow manifold} along which motion obeys $\dot\bx = m_\odot^{-1}(\bu
+ \tau \bu_\tau)$, with an $O(\tau^2)$ error, where $\bu_\tau :=
\big(R\smash{\frac{\D{}}{\D{t}}}\bv + R (f + \tau_\odot v^x +
\frac{1}{3}\omega) \bv^\perp - \smash{\frac{\D{}}{\D{t}}}\bu - (f
+ \tau_\odot v^x + \tfrac{1}{3}R\omega) \bu^\perp\big)$ with
$\smash{\frac{\D{}}{\D{t}}}\bu = \partial_t \bu +
\smash{\gamma_\odot^{-1}(\partial_x \bu)}u^x + (\partial_y \bu)u^y$.
While this result is time-asymptotic, initial behavior can be
anticipated as follows. The slow manifold lies at an $O(\tau)$
distance from a \emph{critical manifold} on which motion obeys
$\dot\bx = m_\odot^{-1}\bu$.  Unlike the slow manifold, the critical
manifold has no global effect on the dynamics of inertial particles
controlled by \eqref{eq:MR}--\eqref{eq:u}.  Yet, due to smooth
dependence of solutions of the BOM equation (1)--(2) on parameters, the
trajectory of a sufficiently small particle will initially run close
to the critical manifold  if $\dot\bx(t_0)$
is close to $(m_\odot^{-1}\bu)(\bx_0,t_0)$, before the particle
starts to drift away from the critical manifold on its way toward
the slow manifold.  This observation is important in practice because
the equations on the critical and slow manifolds do not require one
to specify the velocity at the initial time (they are first-order
equations), which is not known in general.  It is clear, however,
that long-term aspects of the inertial dynamics, such as the dynamics
of great garbage patches \cite{Beron-Vera2016, Beron-Vera2019} or
mesoscale eddies as floating debris traps \cite{Beron-Vera2015,
Beron-Vera2019, Beron-Vera2020}, cannot be described by the equation
on the critical manifold.  These are described by that on the slow
manifold.

\section{Conclusions}

Two main conclusions were reached from the analysis presented in
this letter. First, the trajectories of floating matter are strongly
constrained by their buoyancy.  This has been evidenced quite clearly
from the analysis of the trajectories of custom-made undrogued
drifters with varied designs, which revealed a tendency to cluster
according to buoyancy. In the case of objects such as
pelagic \emph{Sargassum} patches and larger pieces of marine debris,
assuming movement is strictly Lagrangian could have serious
implications in our ability to forecast trajectories and conduct
assessments of observed distribution patterns. This is particularly
important for \emph{Sargassum} as there is considerable uncertainty
associated in determining the importance of localized growth vs
long-distance transport on the inundation events in coastal areas
throughout the Caribbean Sea \cite{Putman2018, Putman2020,
Brooks2019, Johns2020}. The second important conclusion from our
study is that the BOM equation, a recently proposed Maxey--Riley equation
for the motion of finite-size particles floating on the ocean surface
\cite{Beron-Vera2019}, provides a
very reasonable explanation for the observed motion.  We argue that
uncertainties around the ocean current representation (the ocean
velocity considered here did not represent submesoscale aspects of
the motion, such as wave-induced drift) are the main factors
contributing to departures from observed behavior.  This
inference finds additional support on controlled air--water stream
flume  experiments involving spheres of different buoyancies \cite{Miron2020}
that show that motion dependence on inertia characteristics
(particularly drag dependence on buoyancy) is very well described
by the BOM equation. Future studies may
include laboratory experimentation aimed at better framing
arbitrary shaped object motion and incorporating wave-induced drift,
at present represented at the carrying flow level. Field experiments
in a near-coastal area well sampled by high-frequency radars and
weather stations involving actual debris pieces and \emph{Sargassum}
patches would be desirable to test the results of any improvements
to the theory.

\acknowledgments The special drifters of the BOM equation were designed
and constructed by Ulises Rivero and Robert Roddy of the NOAA's
Atlantic Oceanographic and Meteorological Laboratory. The
altimetry/wind/drifter synthesis was produced by RL and can be
obtained from \url{ftp://ftp.aoml.noaa.gov/phod/pub/lumpkin/decomp}.
The ERA-Interim reanalysis is produced by ECMWF and is available
from \url{http://www.ecmwf.int}. This work was supported by the
University of Miami's Cooperative Institute for Marine \& Atmospheric
Studies (PM, MJO, FJBV, and NFP), National Oceanic and Atmospheric
Administration's Atlantic Oceanographic and Meteorological Laboratory
(RL, and GJG), and OceanWatch (JT).


\end{document}